\documentclass[slac]{revtex4}
\usepackage{graphicx,amsmath}
\usepackage{fancyhdr}
\usepackage{slashed,caption}
\usepackage{braket}

\begin{document}

\title{Inverse soft limit construction of QCD amplitude}

%

\author{\textbf{Andriniaina Narindra Rasoanaivo}}

\affiliation{Sciences Exp\'erimentales et des Math\'ematiques, Ecole Normale Sup\'erieure Ampefiloha,\\
Antananarivo - BP 881,  Universit\'e d'Antananarivo, Madagascar\vspace{.2cm}}

\begin{abstract}{
\textbf{\normalsize Abstract:}
QCD amplitudes are one of the most important ingredients for the understanding of the early universe. In this work we present how the knowledge of the asymptotic states can be used to calculate the scattering amplitude of the underline QCD events at high energy. To do so, we show how the understanding of the soft factorization and the soft structure of amplitudes can provide us all the necessary information to fix the kinematics of amplitudes using inverse soft limit techniques.

 }

\end{abstract}

\maketitle

\thispagestyle{fancy}


\section{Introduction}
Quantum field theory is known to be the main framework to study the behaviour of elementary particles and to describe the fundamental interactions between particles. The computation of scattering amplitudes is the central focus to connect such theoretical descriptions to the experimental measurements. The standard method to compute scattering amplitudes is the Feynman diagrammatic approach, however with this technique the computation tends to be very complicated as the number of particles involved in the process get large. In quantum chromodynamic (QCD), the number of gluon tends to grow very fast in the process which makes the computation of amplitudes so challenging to match with the high precision experimental results. 
 
Many progresses have been made toward a more efficient computation, the focus point of such progress is to understand the mathematical structure of scattering amplitudes[]. The standard approach of Feynman is known to make the locality and unitarity structure of the amplitude manifest at every step of the computation. The idea to make both locality and unitarity to manifest at same time is known to be source of all the complexity in the standard computation. A modern approach, like the BCFW recursion of Brito Cachazo Fen Witten \cite{Britto:2004ap}, only make the unitarity manifest along side the on-shell property of the particles involved in the process. Despite the non conventional variables (onshell variables) used in the recursive approach, the computation turns out very efficient to compute a large number point amplitudes. Modern approaches also make various properties and mathematical structure of amplitudes manifest and easy to study.

From our previous work \cite{Rasoanaivo:2022kkm}, we showed how scattering amplitudes can be decomposed around the softness of one of its particle which is an interesting structure around the soft limit theorem \cite{Weinberg:1965nx, Jackiw:1968zza, Elvang:2016qvq}. The two components of such decomposition are first the soft core, which is the main subject to the universal Weinberg soft factorisation, and second the soft shell that needs to be computed in order to fully reconstruct a lager point amplitude from inverse soft limit (ISL) of a lower point amplitudes. Many works have been done toward building full amplitudes from ISL, and such construction can be fully fixed from the asymptotic symmetry of the individual particles involved in process.

In this work, our main objective is to investigate the possibility of construction QCD amplitudes from ISL. We will show from our experiences of soft decomposition of known amplitudes an algorithmic patterns to reconstruct bipolar amplitudes, a configuration class of amplitudes, from ISL. Such algorithm will be presented as graph, oriented graph, showing the different operation to reconstruct full amplitudes.

\section{Inverse soft limit for MHV amplitudes}
Before we talk about the inverse soft limit, it is important to talk about the Weinberg theorem \cite{Weinberg:1965nx}. This theorem shows that a scattering amplitude has an universal factorisation behaviour as the momentum of a massless boson tends to be near zero, \cite{Rasoanaivo:2020yii,Campoleoni:2017mbt}:
\begin{equation}
\lim_{k\to 0} A_{n+1}(p_1,p_2,\ldots,p_n,k)=\hat{S}(k)A_{n}(p_1,p_2,\ldots,p_n),
\label{weinberg}
\end{equation}
where $A_{n}$ is an $n$-point amplitude and $\hat{S}(k)$ is the soft operator related to the soft momentum $k$. This factorisation is known to be universal, and its allow us to connects an $(n+1)$-points amplitude to an $(n)$-point. As shown in \cite{Rasoanaivo:2020yii} the soft operator $\hat{S}$ could be derived independently from the amplitude by resolving the following equation
\begin{equation}
\left [\hat{H}_i,\hat{S}(k_j)\right ]=h_i\hat{S}(k_j)\delta_{ij}. 
\label{soft_constraint}
\end{equation}
In this equation, $\hat{S}(k_j)$ is the soft operator associated to $j$-th particle taken to be soft, $h_i$ is the helicity of the $i$-th particle, $\delta_ij$ is the Kronecker symbol, and $\hat{H}_i$ is the helicity operator associated to $i$-th particle where with the spinor helicity variables \cite{Rasoanaivo:2020yii}
\begin{equation}
\hat{H}_i=-\frac{1}{2}\left (\lambda^a_i\frac{\partial\;}{\partial \lambda^a_i}-\bar{\lambda}^{\dot{a}}_i\frac{\partial\;}{\partial \bar{\lambda}^{\dot{a}}_i}\right ).
\end{equation}

The inverse soft limit is the opposite action where a soft operator is applied on lower point amplitude while its momentum is taken to be hard so the the $n$-point will give rise to an $(n+1)$-point. However as shown in \cite{Rasoanaivo:2022kkm}, such process is not possible with one single ISL. The soft limit can be represented as
\begin{equation}
A_{n+1}\xrightarrow[SL]{\;\text{soft limit}\;}A_n
\end{equation}
while the inverse soft
\begin{equation}
A_n\xrightarrow[ISL]{\;\text{invert soft limit}\;} A_{n+1}^{ISL}=A_{n+1}-R_{n+1},
\end{equation}
where $R_{n+1}$ is the missing part of the higher point amplitude that can be completed from an ISL of other soft sector provided by a different particle. In this soft decomposition, the main objective of the ISL reconstruction is to find an algorithm to calculate $R_n$ so we can complete the amplitude. Around this idea of soft reconstruction, it is worth explore the decomposition of an amplitude around the soft behaviour of a given particle. Let us consider the $i$-th particle in the process, an amplitude can be decompose as the core of the soft theorem of the $i$-th particle $A^{[i]}_n$(the part that can be recovered by ISL), and the soft shell which is lost in this soft limit $R_n^{[i]}$,
\begin{equation}
A_n=A^{[i]}_n+R^{[i]}_n  \Longleftrightarrow \left \{
\begin{aligned}
&A_n\xrightarrow[]{\;k_i\to0\;} A_{n-1}\xrightarrow{\;ISL \:} A_n^{[i]}\\
&R^{[i]}_n\xrightarrow[]{\;k_i\to0\;} 0.
\end{aligned}\right .
\end{equation}

One of the main problem of the standard approach is that amplitudes are computed in a more generalised way where the helicity of particles are only specified at the very last step of the computation. In the modern approaches, especially after Park and Taylor introduce formula for some $n$-points amplitude \cite{Parke:1986gb}, the computation of amplitudes simplified depending on them helicity configuration.

\begin{figure}
\includegraphics[scale=.25]{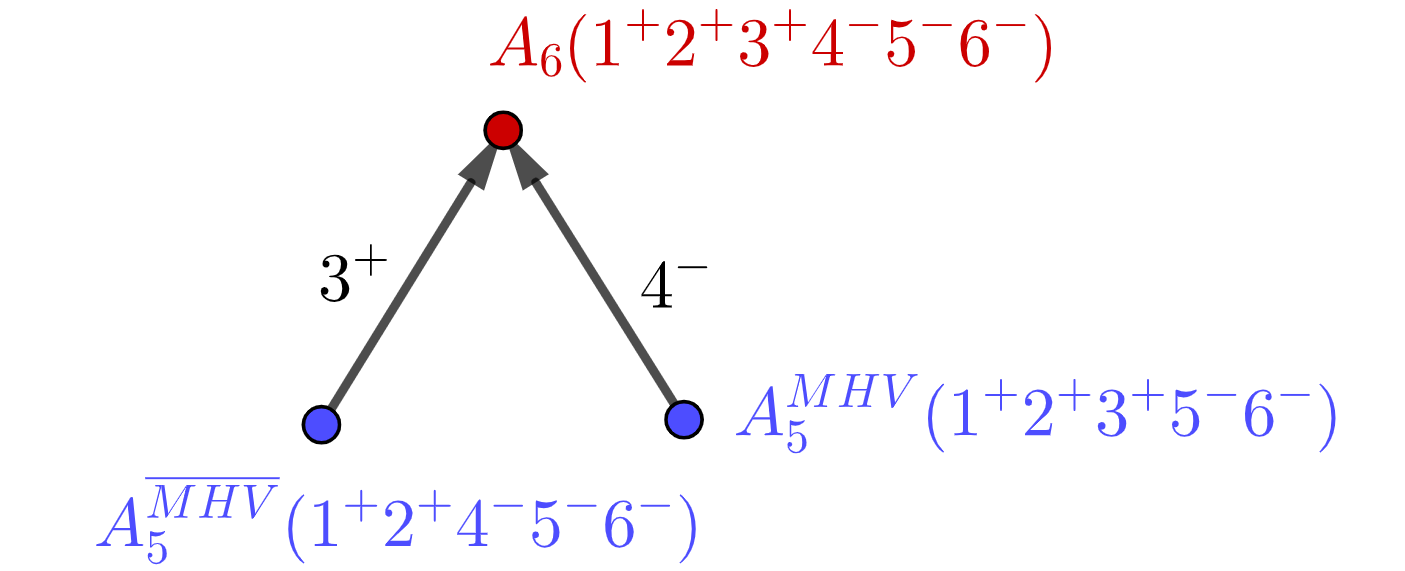}
\caption{Six-point amplitude from one MHV and one $\overline{\text{MHV}}$ amplitudes}
\label{bipolar6}
\end{figure}

In order to understand better these classification let us only consider pure gluon amplitudes were all the particles are all outgoing. Depending on the helicity configuration, for any $n$-point amplitudes, here are the classification of gluon amplitudes
\begin{itemize}
\item \textit{Vanishing amplitudes:} are mplitude vanishes when all the particles have the same helicity or there is only one negative of positive helicity.
\item \textit{MHV amplitudes:} these amplitudes are the first non vanishing amplitudes in most non conserved way of the helicity in which we only have two negative helicity particles.
\item \textit{$\overline{\text{MHV}}$ amplitudes:} the MHV bar amplitudes are similar to the MHV  amplitudes in which we only have two positive helicity particles.
\item \textit{N${}^k\!$MHV amplitudes:} are amplitudes that have $k+2$ negative helicity particles, $k$-th next the MHV amplitudes. 
\end{itemize}

The well understood among these amplitudes are the MHV and the $\overline{\text{MHV}}$ amplitudes. Them expression can be derived directly from the Park-Taylor formula \cite{Parke:1986gb}. And the amazing part is that in a soft limit where $A_{n+1}^{MHV}$ amplitudes is factorized into a soft operator and a MHV sub-amplitude $A_n^{MHV}$ a single inverse soft limit is enough to recover the original amplitude,
\begin{equation}
A_{n+1}^{MHV}\xrightarrow{\; \text{ soft limit }\;} A_n^{MHV}\xrightarrow{\; \text{ inverse soft limit }\;}A_{n+1}^{MHV}.
\end{equation}
This stability of the MHV and $\overline{\text{MHV}}$ amplitudes and them simplicity are the main reasons we choose MHV amplitude to be central part of amplitudes. For example the 6-point amplitude $A_6(1^-2^-3^-4^+5^+6^+)$ [2205.13873], it can be constructed for two inverse soft limits as shown by the graph in the Fig. \ref{bipolar6}.

\section{Bipolar amplitudes}

\begin{figure}
\includegraphics[scale=.25]{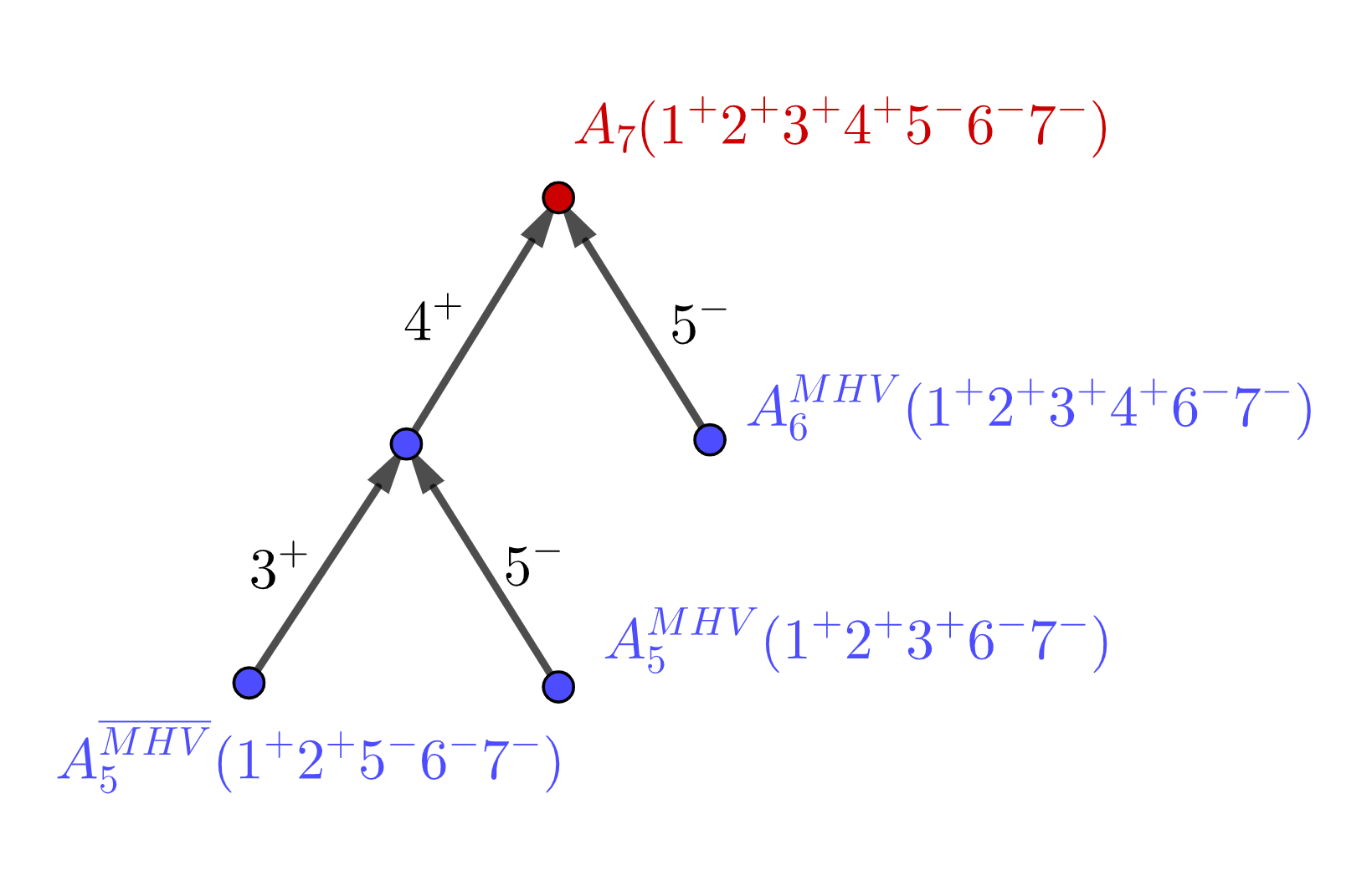}
\caption{Seven-point bipolar amplitude reconstruction.}
\label{bipolar7}
\end{figure}

In our experiments, we analysed the soft decomposition of many different helicity amplitudes. From these decomposition, we explore the possibility of inverting the process using ISL calculations. It is important to mention that we are in the search of pattern to generalise the ISL reconstruction of any point amplitude.

Out of these experiments, we found that there seems a pattern in some class of helicity configurations. These amplitude have all of its positive helicity grouped in one region. A typical form of that is when all the positive helicity is grouped to the right and all the negative to left such as $A_6(1^-2^-3^+4^+5^+)$, $A_6(1^-2^-3^-4^+5^+6^+)$, $A_6(1^-2^-3^-4^+5^+6^+7^+)$, \ldots  Here we can see that a bipolar amplitude can ever MHV or any of the N$^k$MHV amplitudes since we do not care about the number the negative helicity but more on how are they grouped as bipolar objects.

\begin{figure}
\includegraphics[scale=.25]{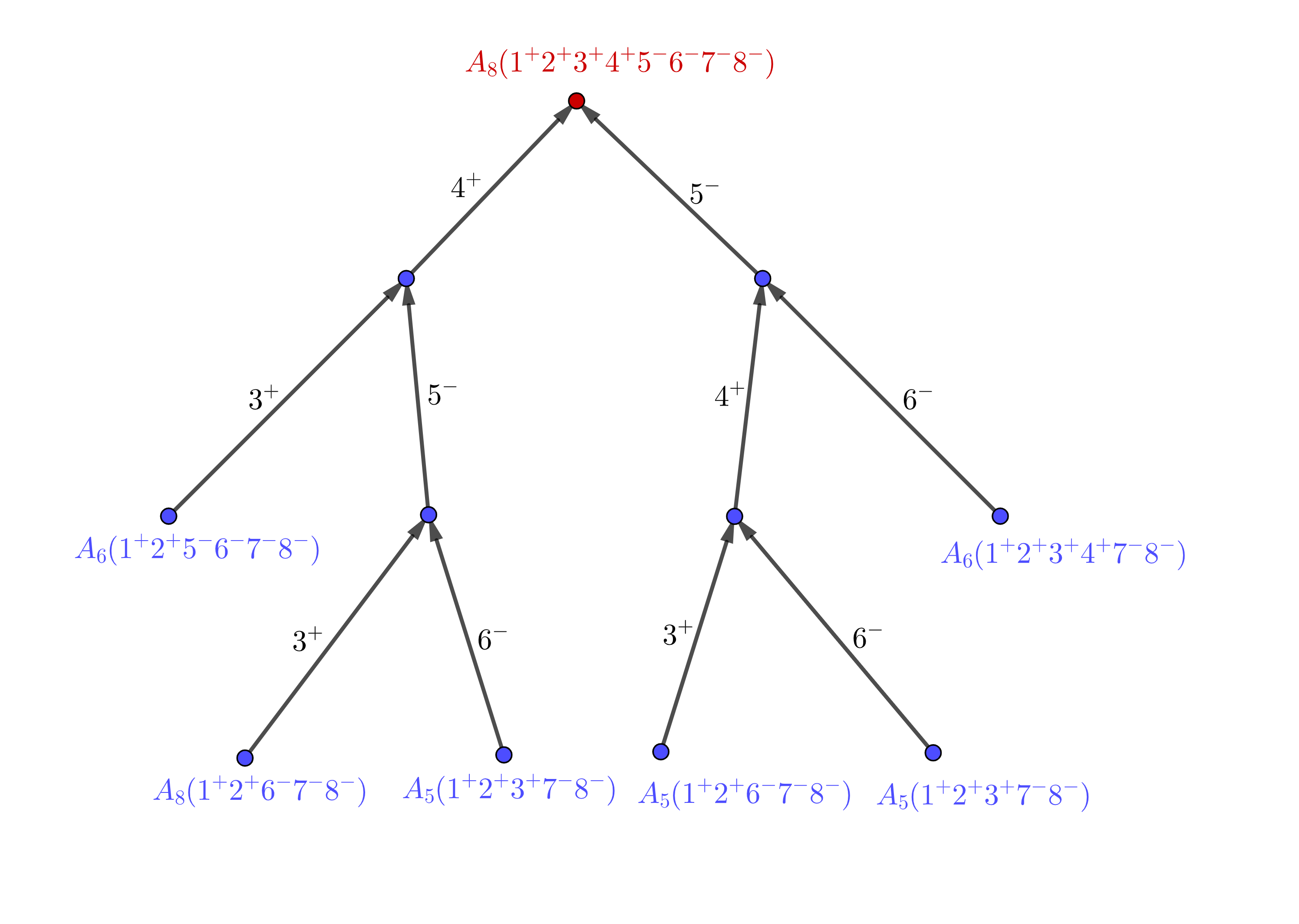}
\caption{Eight-point bipolar amplitude reconstruction.}
\label{bipolar8}
\end{figure}

For simplicity we are calling these amplitude bipolar amplitude. In the soft decomposition experiments we found that any $n$-point  bipolar amplitude can be constructed directly from two ISL of some other  $(n-1)$-points bipolar. This finding allow to generate an ISL algorithm to reconstruct those amplitude from MHV and/or $\overline{\text{MHV}}$ amplitudes. The algorithm can be represented in a bi-parted graph in which:
\begin{itemize}
\item \textit{the root:} is the bipolar amplitude that we aim to reconstruct.
\item \textit{the leaves:} are MHV/$\overline{\text{MHV}}$ amplitudes which are the seeds of the construction.
\item \textit{the oriented edges:} are the ISL operations that connect the $n$-points to the $(n+1)$-points.
\item \textit{the intermediate vertices:} are the intermediate bipolar amplitude that connects the MHV's to the root.
\end{itemize}
The Fig. \ref{bipolar7} and Fig. \ref{bipolar8} are two examples of the calculation we did using the bi-parted algorithm above. It is worth to mention that the ISL operation made at every edge in the graph consist to multiply the amplitude with the corresponding soft operator of the particle labelled on the arrow and to use a momentum shift as presented in \cite{Boucher-Veronneau:2011rwd, Nandan:2012rk}

\section{Conclusion}

In conclusion, we aim to construct QCD amplitudes from the asymptotic behaviour of the individual particles using inverse soft limit. And we found that such process is well establish for the case of MHV amplitudes since they have a simplicity in the soft decomposition. Such decomposition is main ingredient of this reconstruction, and the analysis of decomposition of many different amplitudes allow us to observe a pattern in the soft structure certain amplitudes we called bipolar amplitudes. In the present work presented at the \textit{HEPMAD22}, we only gave the ISL algorithm, the formal proof will be presented in a separate paper of our coming works. The extension of the method can be considered for any helicity configuration, the only complication of some helicity configuration is the apparition of loops in the graph associated to the ISL algorithm.

%
%
%
%

\end{document}